\documentclass[%
 amsmath,amssymb,
 aps,
11pt,
]{revtex4-2}

\usepackage[all]{background} 
\usepackage{url}
\SetBgContents{Published by Physics of Fluids}
\SetBgColor{gray}
\SetBgPosition{5.5,-16}
\SetBgOpacity{0.75}
\SetBgAngle{0.0}
\SetBgScale{1.5}

\usepackage{tikz,xcolor,hyperref}
\definecolor{lime}{HTML}{A6CE39}
\DeclareRobustCommand{\orcidicon}{%
	\begin{tikzpicture}
	\draw[lime, fill=lime] (0,0) 
	circle [radius=0.16] 
	node[white] {{\fontfamily{qag}\selectfont \tiny ID}};
	\draw[white, fill=white] (-0.0625,0.095) 
	circle [radius=0.007];
	\end{tikzpicture}
	\hspace{-2mm}
}
\foreach \x in {A, ..., Z}{%
	\expandafter\xdef\csname orcid\x\endcsname{\noexpand\href{https://orcid.org/\csname orcidauthor\x\endcsname}{\noexpand\orcidicon}}
}

\usepackage{graphicx}
\usepackage{dcolumn}
\usepackage{bm}

\newcommand{\bsub}{\begin{subequations}}
	\newcommand{\esub}{\end{subequations}}

\newcommand{\be}{\beta}

\newcommand{\om}{\omega}

\newcommand{\de}{\delta}

\newcommand{\Ga}{\Gamma}

\newcommand{\po}{\mbox{\boldmath $\omega$}}

\newcommand{\pt}{\mbox{\boldmath $\tau$}}

\newcommand{\ppsi}{\mbox{\boldmath $\psi$}}

\newcommand{\pGamma}{\mbox{\boldmath $\Gamma$}}

\newcommand{\bG}{\mathbf  G}

\newcommand{\bI}{\mathbf  I}

\newcommand{\bM}{\mathbf  M}

\newcommand{\bT}{\mathbf  T}

\newcommand{\pD}{\textbf{\emph{D}}}

\newcommand{\pF}{\textbf{\emph{F}}}

\newcommand{\pL}{\textbf{\emph{L}}}

\newcommand{\pU}{\textbf{\emph{U}}}

\newcommand{\pe}{\textbf{\emph{e}}}
\newcommand{\pf}{\textbf{\emph{f}}}

\newcommand{\pn}{\textbf{\emph{n}}}

\newcommand{\pr}{\textbf{\emph{r}}}

\newcommand{\pu}{\textbf{\emph{u}}}
\newcommand{\pv}{\textbf{\emph{v}}}

\newcommand{\px}{\textbf{\emph{x}}}

\newcommand{\ptt}{\textbf{\emph{t}}}

\newcommand{\pat}{\partial}
\newcommand{\na}{\nabla}
\newcommand{\x}{\times}

\newcommand{\beq}{\begin{equation}}
\newcommand{\eeq}{\end{equation}}
\newcommand{\bsubeq}{\begin{subequations}}
	\newcommand{\esubeq}{\end{subequations}}
\newcommand{\beqn}{\begin{eqnarray}}
\newcommand{\eeqn}{\end{eqnarray}}
\newcommand{\fr}{\frac}
\newcommand{\lb}{\label}
\newcommand{\er}{\eqref}

\usepackage{hyperref}
\hypersetup{colorlinks,allcolors=blue}
\usepackage{amsmath}
\usepackage{multirow}
\usepackage{amsfonts,amssymb}
\usepackage{graphicx}
\usepackage{subfigure}
\usepackage[rel]{overpic}

\def\Xint#1{\mathchoice
{\XXint\displaystyle\textstyle{#1}}%
{\XXint\textstyle\scriptstyle{#1}}%
{\XXint\scriptstyle\scriptscriptstyle{#1}}%
{\XXint\scriptscriptstyle\scriptscriptstyle{#1}}%
\!\int}
\def\XXint#1#2#3{{\setbox0=\hbox{$#1{#2#3}{\int}$}
\vcenter{\hbox{$#2#3$}}\kern-.5\wd0}}

\def\dashint{\Xint-}

\graphicspath{{figures/}}

\begin{document}

\preprint{APS/123-QED}


\title{Lift and drag in three-dimensional steady viscous and compressible flow}

\author{Luoqin Liu\orcidA{}}
 \email{lqliu@pku.edu.cn}

\author{Jiezhi Wu}

\author{Weidong Su}

\author{Linlin Kang}

\affiliation{State Key Laboratory for Turbulence and Complex Systems, Center for Applied Physics and Technology, College of Engineering, Peking University, Beijing 100871, China}

\date{June 12, 2017}

\begin{abstract}
In a recent paper, Liu, Zhu and Wu (2015, {\it J. Fluid Mech.} {\bf 784}: 304) present a force theory for a body in a two-dimensional, viscous, compressible and steady flow. In this companion paper we do the same for three-dimensional flow. Using the fundamental solution of the linearized Navier-Stokes equations, we improve the force formula for incompressible flow originally derived by Goldstein in 1931 and summarized by Milne-Thomson in 1968, both being far from complete, to its perfect final form, which is further proved to be universally true from subsonic to supersonic flows. We call this result the \textit{unified force theorem}, which states that the forces are always determined by the vector circulation $\pGamma_\phi$ of longitudinal velocity and the scalar inflow $Q_\psi$ of transverse velocity. Since this theorem is not directly observable either experimentally or computationally, a testable version is also derived, which, however, holds only in the linear far field. We name this version the \textit{testable unified force formula}. After that, a general principle to increase the lift-drag ratio is proposed.
\end{abstract}


\maketitle

\section{Introduction}\label{sec.Introduction}

The object of this paper is to obtain formulae for the forces when a solid body of any shape is at rest in a steady stream of viscous and compressible fluid. Namely, we assume the body moves steadily through the physical space filled with {restful} fluid, let the reference frame fix on the body, and neglect the inherent flow unsteadiness at very far field \citep{Liu2017}. In such a flow model we only work in the steady subspace $V_{\rm st}$ of the whole physical space, which can be divided into a nonlinear near-field and a linear far-field. Correspondingly, the aerodynamic force theories can also be categorized into two groups. One is the \textit{far-field force theories}, by which universal force formulae can be rigorously deduced. Its central task is to identify the key physical quantities responsible for the forces, of which the first and most classic example is the Kutta-Joukowski (K-J) theorem \citep{Joukowski1906} for inviscid and incompressible flow that focuses one's attention to the circulation around an airfoil. The other is the {\it near-field theories}, which can determine the relationships between the forces and the detailed flow processes and structures behind those universally identified key quantities. It is guided (for steady flow) by the results of far-field theories and always the main body of aerodynamic theories, as given in all monographs and textbooks of low- and high-speed aerodynamics.

Historically, however, various far-field theories had long been limited to incompressible and/or inviscid flow and never reached their highest possible goal to be truly universal: to identify the key physical quantities responsible for aerodynamic forces within the general framework of the Navier-Stokes (N-S) equations. The first breakthrough was made only very recently by \citet[LZW for short]{Liu2015}, who obtained a unified far-field aerodynamic force theory for two-dimensional (2D) viscous and compressible flow, valid from low-speed to supersonic regimes. In this theoretical paper, we present an exactly the same kind of theory but for three-dimensional (3D) flow.

To explain the motivation and orientation of our study, it is appropriate here to make a brief account of previous investigations about the far-field steady force theories, for both 2D and 3D flows due to their close relations. In so doing we rely crucially on the  so-called \textit{Helmholtz decomposition} \citep{Stokes1851, Helmholtz1858}, which decomposes a vector field into a longitudinal part and a transverse part. Let $\pu$ and $\pU$ denote the local and incoming flow velocities, then the disturbance velocity $\pu'=\pu - \pU$ can always be written formally as
\beq\lb{HD}
  \pu'=  \pu_\phi+\pu_\psi \equiv\na\phi+\na\times \ppsi, \quad \na\cdot\ppsi=0,
\eeq
where $\phi$ and $\ppsi$ are the \textit{velocity potential} (scalar potential) of the longitudinal field and the \textit{vortical stream function} (vector potential) of the transverse field, respectively.

Before proceeding, we remark that the existence and uniqueness of the Helmholtz decomposition \er{HD} is not always ensured, which is strongly dependent on the boundary conditions as well as the flow domains. For the uniqueness problem, \citet[pp.~37-38]{Chorin2000} have obtained a positive answer under the assumption that the transverse velocity $\pu_\psi$ is parallel to the boundary of domain. This powerful theorem, however, is not applicable to our case since in general the parallelism assumption can not be satisfied at the body surface \citep[pp.~43-44]{Wu2015}. Fortunately, for the linearized flow, which is the concern of present paper, \citet[pp.~30-34]{Lagerstrom1949} have proved that the decomposition \er{HD} is indeed unique such that the N-S equation can be thoroughly split into two parts (see \er{NS} below). For the existence problem, \citet[pp.~164-165]{Serrin1959} has shown that it can be satisfied for both finite domain with the normal vorticity being zero at the boundary and infinite domain with the velocity decaying fast enough at far field. For the latter case, \citet{Gregory1996} has proved that the result is still valid even if isolated singular points present. Therefore, the existence and uniqueness of \er{HD} in the case of LZW is completely confirmed since the whole space is regular except the isolated singular point where the body locates. But Gregory's proof is not applicable to our case as the result (see \er{Psi-2} below) shows that the positive $x$-axis itself (rather than isolated points) is singular. In this sense, further proofs of the existence of \er{HD} under more general conditions are still needed, which is out of the scope of this paper. Nevertheless, the existence of \er{HD} can be taken for granted as long as we can find the corresponding solution under specific boundary conditions \citep[say, far-field decaying condition,][]{Liu2017}.

\subsection{Far-field force theory in two dimensions}

\begin{figure}
  \centering
  \includegraphics[width=0.7\textwidth]{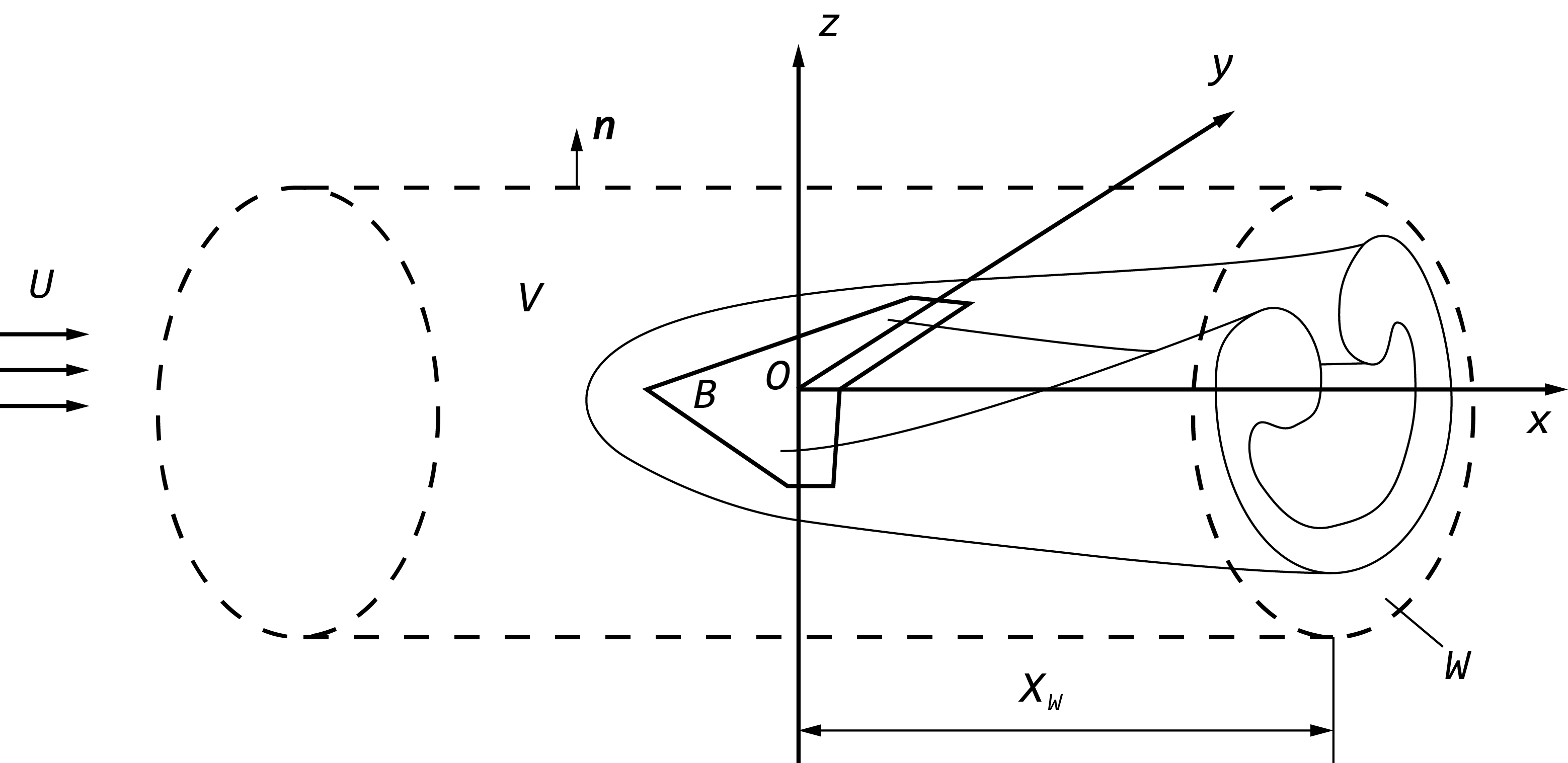}
  \caption{Cartesian coordinates fixed on the body and other notations used in this article}
  \label{coordinate}
\end{figure}

To make the notations unified with 3D flow in Cartesian coordinates $(x,y,z)$ with $\pe_x, \pe_y$ and $\pe_z$ being the unit vectors along incoming-flow, spanwise and vertical-up directions, respectively (see Fig.~\ref{coordinate}), a 2D flow is assumed to occur on a $(x,z)$-sectional plane, with vorticity $\po=\om\pe_y$ and $\ppsi =(0,\psi,0)$ so that $\pu_\psi =\na\psi\times \pe_y$. A control surface $S$ with unit normal vector $\pn$ in three dimensions {\color{black} reduces} to a closed loop, which has unit tangent vector $\ptt$ so that $\pn\times \ptt =\pe_y$. Then the lift $\pL=L\pe_z$ on a solid body in a steady flow of incompressible and inviscid fluid reads
\beq\lb{KJ}
  \pL =\rho_0 \pU \times\pGamma_\phi,
\eeq
where $\rho_0$ and $\pU = U \pe_x$ are the fluid density and velocity, respectively, at infinity,
\begin{equation}\label{G-phi}
 \pGamma_\phi \equiv \int_S \pn \times \na \phi \mathrm{d} S = \pe_y[\![\phi]\!]
\end{equation}
is the vector circulation $\pGamma =\Ga\pe_y$ of a bound vortex in the body, $\phi$ is the velocity potential and $[\![\cdot]\!]$ denotes the jump of the corresponding quantity as one goes around the loop once. Equation \er{KJ} is the well-known K-J theorem, which is completely independent of the size and geometry of $S$ and has since served as the very basis of classic steady aerodynamics. For example, when the body is an airfoil with sharp trailing edge, $\pGamma_\phi$ can be determined by the Kutta condition \citep{Kutta1902}. Meanwhile, for this inviscid flow the drag is zero, in consistent with the famous \textit{d'Alembert paradox}.

In contrast to inviscid flow, in a viscous flow the vortical wake must extend to downstream {\color{black} all the way to the outside of the steady flow region $V_{\rm st}$ so that any contour $S \in V_{\rm st}$} surrounding the airfoil must cut through the wake, leaving some vorticity outside of $S$. Thus, one has to ask (i) whether \er{KJ} is still effective, and {\color{black} if yes, (ii)} whether the lift is still independent of the choice of $S$.

These questions were first studied experimentally by \citet{Bryant1926}. They found that the lift calculated by \er{KJ}, with $\na\phi$ replaced by $\pu'$ {\color{black} or $\pu$}, is a good approximation to that of the real viscous flow for typical aerodynamic applications:
\beq\lb{Taylor-L}
  \pL \cong \rho_0 \pU \times\pGamma,
\eeq
{\color{black} with $\cong$ implying \textit{asymptotic equality} throughout this paper and}
\begin{equation}\label{Gamma}
 \pGamma \equiv \int_S \pn \times \pu' \mathrm{d} S= \int_V\po \mathrm{d} V.
\end{equation}
{\color{black} Here $\po = \na\times \pu'$ is the vorticity and} $V$ is the volume enclosed by $S$. The experiment {\color{black} results also indicate} that $\pGamma$ may still be independent on $S$. {\color{black} Then}, \citet{Taylor1926} pointed out that these {\color{black} two} positive answers require two conditions:  (a) the intersect of $S$ and the wake has to be a vertical plane (``wake plane'', denoted by $W$) with normal $\pn =\pe_x$; and (b) the net vorticity flux through $W$ must vanish, which can be proven for steady flow at large Reynolds number {\color{black} $Re$} (for an improved proof of this issue see Ref.~\onlinecite{Wu2015}). We call these conditions the {\it first and second Taylor criteria} \citep{Liu2015}, and \er{Taylor-L} the Taylor lift formula.

Independent of the work of \citet{Taylor1926}, \citet{Filon1926} made a thorough analysis of the lift and drag problem for 2D {\color{black} incompressible and viscous} flow. He {\color{black} confirmed} that to the leading order the disturbance flow satisfies the Oseen equation (see \er{u-t} below), which is valid for an arbitrarily {\color{black} $Re$} as long as the distance from the body is sufficiently large. After obtaining the complete solution of Oseen's equation, {\color{black} he finally} showed that the lift is the same as \er{Taylor-L} at infinity,
while the drag is associated with a particular term in the solution, given by
\begin{equation}\label{D-3D}
 \pD = \rho_0 \pU Q_\psi,
\end{equation}
where
\begin{equation}\label{Q-psi}
 Q_\psi \equiv -\int_S (\pn \times \na) \cdot \ppsi \mathrm{d} S =-[\![\psi]\!]
\end{equation}
represents an {\it inflow} at infinity at the tail.

Now, what LZW has achieved is to extend the above lift and drag formulae {\color{black} of incompressible flow to compressible and viscous flow.} Starting from \er{HD} and to make the decomposition unique, the linearized compressible N-S equation has to be split into a longitudinal {\color{black} part and a transverse part \citep[pp.~30-34]{Lagerstrom1949},}
\bsubeq\lb{NS}
\beqn
 \Pi + \rho_0 U \fr{\pat \phi}{\pat x}  &=& 0, \lb{u-l} \\
 \left( \na^2 - 2k \fr{\pat}{\pat x} \right) \pu_\psi &=& \bf 0, \lb{u-t}
\eeqn\esubeq
where $\Pi =p-\mu_\theta \vartheta$ is the {\color{black} modified pressure, $\vartheta = \na \cdot \pu$ is the dilatation, $k = \rho_0 U/2\mu$, and $\mu_\theta$ and $\mu$ are the longitudinal and shear viscosities, respectively.} While \er{u-l} is the linearized Bernoulli equation, \er{u-t} is the transverse Oseen equation. Then, {\color{black} using} the far-field analysis and the fundamental solution method, LZW {\color{black} has} proven that the force $\pF$ exerted on the body is
\beq\lb{F-2D}
 \pF =\rho_0\pU\times \pGamma_\phi +\rho_0 \pU Q_\psi,
\eeq
where $\pGamma_\phi$ and $Q_\psi$ are given by \er{G-phi} and \er{Q-psi}, respectively. Obviously, \er{F-2D} is independent of the choice of $S$ since both \er{G-phi} and \er{Q-psi} are independent of $S$ due to the generalized Stokes theorem \citep[e.g.,][p.~700]{Wu2006}. Since no assumption of incompressibility {\color{black} is used} in the derivation of \er{F-2D}, it is also true for compressible flow in a wide range of Mach number, from subsonic to supersonic flows. Due to this {\color{black} universality}, LZW calls \er{F-2D} the Joukowski-Filon {\color{black} (J-F)} theorem, which states that the lift and drag are always determined by $\pGamma_\phi$ and $Q_\psi$, {\color{black} respectively,} no matter how complicated the near-field flow surrounding the body might be.

Unfortunately, velocity potentials {\color{black} $\phi$ and $\ppsi$} are not directly observable either experimentally or computationally, and hence neither are the integrands of the J-F theorem. This is why Filon's drag formula has seldom been noticed in aerodynamics community. But except providing universal and exact force {\color{black} formulae}, the far-field theories have another task, namely to give asymptotic {\color{black} formulae that are expressed by observable variables} only. Thus, LZW also derived a testable version of the J-F formula:
\begin{equation}\label{F-2D-2}
 \pF \cong \rho_0 \pU \times \pGamma + \rho_0 \pU Q_W,
\end{equation}
where $\pGamma$ is given by \er{Gamma}, {\color{black} $W$ denotes the wake plane} and
\begin{equation}\label{Q-W}
 Q_W \equiv \int_W  z \om \mathrm{d} S.
\end{equation}
{\color{black} Evidently,} $\pGamma$ and $Q_W$ depend linearly on the vorticity and thus \er{F-2D-2} {\color{black} should still hold} in the linear far field of time-averaged {\color{black} steady} flow such as turbulence, {\color{black} which has been confirmed by numerical studies (cf.~LZW)}.

Owing to this progress, the far-field force theory has for the first time been rigorously extended to {\color{black} compressible and viscous} flow. The J-F theorem is thus far the only force theory that has the same form in incompressible and compressible flows. {\color{black} In this sense}, the far-field force theory in two dimensions {\color{black} is complete}.

\subsection{Previous works on far-field force theory in three dimensions}

The corresponding incompressible problem in three dimensions has been treated in two papers by \citet{Goldstein1929, Goldstein1931}, who followed \citet{Filon1926} to apply the Oseen approximation at a great distance from the solid. In his {\color{black} 1929} paper, Goldstein discussed two series of solutions of the {\color{black} velocity field}, which corresponds exactly to {\color{black} $\pu_\phi$ and $\pu_\psi$} given by \er{HD}. The first series yields a set of particular integrals of \er{u-l}, {\color{black} where} the velocities are associated with certain values of the pressure. In the second series, which is a set of particular integrals of \er{u-t} and of the nature of a complementary function, the velocities are rotational while the pressure does not appear. Thus, Filon's drag formula \er{D-3D} is shown to still {\color{black} hold. However,} as pointed out by Burgers in one of notes of \citet{Goldstein1931}, who was thus listed as a coauthor, the solution of \citet{Goldstein1929} is valid only when the solid body is of revolution, and thus they are not sufficiently general.

In his 1931 paper, Goldstein investigated some more particular integrals, with special emphasis on singular solutions. He showed that for certain values of the pressure, the {\color{black} longitudinal velocity $\pu_\phi$ has} singularities, which have to be cancelled by the suitable component of the {\color{black} transverse} velocity $\pu_\psi$. {\color{black} Furthermore}, he divided $\pu_\psi$ into three parts, $\pv_1, \pv_2$ and $\pv_3$, each of which satisfies \er{u-t}. Then, $\pv_1$ cancels out the singularities in {\color{black} $\pu_\phi$}, the sum $\pv_1+\pv_2$ satisfies the condition of continuity $\na\cdot(\pv_1+\pv_2)=0$, and $\pv_3$ satisfies {\color{black} both the continuity condition and the far-field decaying condition}. Based on these analyses, \citet{Goldstein1931} obtained three major results:

(i) Filon's drag \er{D-3D} is found still true in three dimensions.

(ii) {\color{black} A simple expression for the lift at infinity} is obtained, which is the same as Taylor's lift formula \er{Taylor-L}.

(iii) {\color{black} The} force $\pF$ can be expressed as
\begin{equation}\label{F-3D}
  \pF \cong - \rho_0 U \int_W (\pv_2+\pv_3) \mathrm{d} S,
\end{equation}
so that $D$ is the integral of $\rho_0 \pU\cdot (\pv_2+\pv_3)$ over $W$.

Subsequently, \citet{Garstang1936} obtained the complete {\color{black} solutions} discussed by \citet{Goldstein1931}, and thereby proved $\pv_2\cdot\pU = 0$. {\color{black} Besides}, in describing the results of \citet{Goldstein1931} and \citet{Garstang1936}, \citet[pp.~702-706]{Thomson1968} found that $\pGamma$ defined by \er{Gamma} can be further reduced to the circulation solely due to $\pv_2$. {\color{black} Namely,} in \er{Taylor-L} and \er{F-3D} one may set
\beq\lb{Milne-LD}
\pGamma \cong\int_S\pn\times \pv_2 \mathrm{d} S,\quad D\cong -\rho_0\int_W\pU\cdot \pv_3 \mathrm{d} S.
\eeq

{\color{black} In sharp contrast to 2D} flow, the above results have not yet been pursued to a mature {\color{black} stage: (i)} There is no universal force formula yet except Filon's formula \er{D-3D}; {\color{black} (ii) The force formulae \er{Taylor-L} and \er{F-3D}} are valid only {\color{black} asymptotically; (iii)} The separate appearance of $\pv_2$ and $\pv_3$ in lift and drag of \er{Milne-LD}, respectively, is physically quite strange. {\color{black} In our view,} the main reason for this embarrassing situation is that these authors did not thoroughly utilize the process decoupling \er{HD} and \er{NS}, as seen from their division of $\pu'$ into four parts. Besides, nor did they find a simple method for solving far-field equations. Thus we may ask: if we persist the process decoupling and turn to the fundamental-solution method that has been proven for 2D flow to be much neater and more straightforward than those classic techniques, is there any force formula in three dimensions that is as neat and universal as the {\color{black} J-F theorem \er{F-2D}?} In this paper, we will give a positive answer to this question.

Naturally, the next relevant extension would be compressible flow. Toward this goal and within three dimensions, among others, \citet{Finn1957} have proven rigorously for subsonic nonlinear potential flow that the fluid exerts no net force to the body, which may be termed a ``d'Alembert-like paradox'' but is of course not the case for viscous flow. Then, based on some plausible assumptions, \citet[pp.~34-38]{Lagerstrom1964} has proposed that \er{Taylor-L} should give the lift for viscous compressible flow.

Having reviewed these pieces of progress of 3D far-field analyses, we may conclude that so far no profound and universal force theory is available. In other words, the far-field force theory in three dimensions is still far from complete.

\subsection{Our work and this paper}

In the rest of this paper we extend {\color{black} LZW's theory} to three dimensions in the same way. {\color{black} In Section \ref{sec.far-field} {\color{black} we observe that} the velocity potentials $\phi$ and $\ppsi$ {\color{black} must} be singular}, for otherwise the body would be force-free. We then confirm the effectiveness {\color{black} of Taylor's} lift formula \er{Taylor-L}, as well as a drag formula, {\color{black} for 3D viscous and compressible flow}. In Section \ref{sec.GG} we introduce the fundamental solutions of the linearized N-S equations and make a detailed analysis of the {\color{black} transverse} far-field {\color{black} so that the singularity of $\phi$ and $\ppsi$ is proved rigorously.} Then we arrive at a profound universal force formula, which we state as the \textit{unified force theorem} (UF theorem for short). This neat theorem is however not yet a complete aerodynamic theory since it is not directly testable or measurable. {\color{black} In Section \ref{sec.TGG}, therefore, we propose a type of testable UF formula, which is expressed solely by observable variables.} In Section \ref{sec.Physical} we provide a simple physical explanation of the singularity in incompressible flow with a concrete model, which is the essence of the universality of the UF theorem. A basic principle to increase the lift-drag ratio of {\color{black} 2D and 3D flows} is proposed in Section \ref{sec.L-D}. {\color{black} Conclusions} are given in Section \ref{sec.Conclusions}.

To be self-contained, the fundamental solution of 3D steady linearized N-S equations is given in Appendix \ref{appA} {\color{black} and some details in proving the UF theorem are listed in Appendix \ref{appB}.}

\section{Far-field force formulae and their implications}\label{sec.far-field}

For {\color{black} steady flow}, the total force $\color{black} \pF$ exerted on the body $B$ {\color{black} is}
\bsubeq
\begin{eqnarray}
  \pF &\equiv& -\int_{\pat B} (-\Pi \pn + \pt) \mathrm{d} S \label{11.1a}\\
   &=& -\int_S (\Pi \pn +\rho \pu \pu\cdot \pn-\pt) \mathrm{d} S, \label{11.1b}
\end{eqnarray}
\esubeq
where $\pat B$ is the boundary of body, $S$ is an arbitrary control surface enclosing the body, {\color{black} $\Pi$ is the modified pressure, $\rho$ is the fluid density} and $\pt = \mu \po \times\pn$ is the shear stress {\color{black} with $\mu$ being the constant shear viscosity}.

Hereafter we assume $S$ lies in sufficiently far {\color{black} away} where the flow can be linearized and is governed by \er{NS}. {\color{black} Using} the exact continuity equation $\na\cdot (\rho \pu)=0$ and omitting higher-order terms, there is
\beq
 \int_S \rho\pu\pu\cdot\pn \mathrm{d} S = \int_S \rho_0(\na\phi+\pu_\psi)\pU\cdot\pn \mathrm{d} S,
\eeq
{\color{black} where $\rho_0$ and $\pU$ are the density and velocity of the fluid at infinity, and the Helmholtz decomposition \er{HD} has been used.}
Thus, the linearized version of \er{11.1b} is
\beqn
  \pF &=& \rho_0 \pU\times \int_S \pn\times \na\phi \mathrm{d} S -\rho_0 \pU\cdot \int_S \pn \pu_\psi \mathrm{d} S +\mu \int_S \po\times \pn \mathrm{d} S,\lb{F1}
\eeqn
where the longitudinal equation \er{u-l} has been used. Then, to transform the shear stress $\mu\po\times \pn$ we use the transverse equation \er{u-t}. Since
\begin{equation*}
  \na \times (\pU\times \pu_\psi) = \pU\na\cdot \pu_\psi - \pU\cdot \na \pu_\psi = -\pU\cdot \na \pu_\psi,
\end{equation*}
\er{u-t} can be recast to
\begin{equation*}
  \na\times (\pU\times \pu_\psi)=\na\times (\nu\po), \color{black} \quad
  \nu = \mu/\rho_0,
\end{equation*}
so that
\beq\lb{eta}
  \pU\times \pu_\psi =\nu\po + \na \eta
\eeq
for some scalar function $\eta$, which satisfies the Poisson equation
\beq\lb{eta1}
  \na^2 \eta = - \pU \cdot \po.
\eeq
Thus, from $\na^2 \ppsi = -\po$ follows $\eta= \pU \cdot \ppsi$, and we have
\beq\lb{shear2D}
  \nu\po\times \pn = \pu_\psi\pn\cdot \pU-\pU\pn\cdot \pu_\psi +\pn \times \na (\pU \cdot \ppsi).
\eeq
Then, substituting \er{shear2D} into \er{F1} yields immediately
\beq\lb{F2D}
 \pF = \rho_0\pU\times \pGamma_\phi + \rho_0\pU Q_\psi + \rho_0 \int_S \pn \times \na (\pU \cdot \ppsi) \mathrm{d} S,
\eeq
where {\color{black} $\pGamma_\phi$ and $Q_\psi$} are defined by \er{G-phi} and \er{Q-psi}, respectively. {\color{black} Remarkably, in two dimensions the third term of \er{F2D} vanishes since $\pU\cdot \ppsi = 0$, so that the J-F formula \er{F-2D} recovers. In three dimensions, however, this term} can also contribute to a lift via its $z$-component, which is directly associated with the vortical stream function $\ppsi$.

{\color{black} According} to the generalized Stokes theorem \citep[p.~700]{Wu2006} \er{F2D} would be identically zero unless $\phi$ {\color{black} or} $\ppsi$ are either multi-valued or singular. This general ``d'Alembert-like paradox'' extends {\color{black} what} observed by \citet{Finn1957} to {\color{black} viscous and compressible} flow. For {\color{black} steady flow} over a body, therefore, {\sl it is only the singularity or  multi-valueness of velocity potentials {\color{black} $\phi$ and $\ppsi$} that can ensure nonzero forces}. Indeed, in two dimensions $\phi$ and $\ppsi$ must be multi-valued as discussed by LZW, while in a 3D singly-connected domain $\phi$ and $\ppsi$ have to be singular, as first pointed out by \citet{Goldstein1931} for incompressible flow. {\color{black} We shall see later that this singularity does exist even for compressible flow.}

Of course, \er{F2D} is not yet the final form of force theory. But it can hardly be further pursued without knowing the specific singular property of $\phi$ and $\ppsi$. This will be done in the next section with the help of fundamental solutions, where the last term in \er{F2D} will be proven to be exactly equal to the first term. Consequently, once again, {\color{black} the forces} are totally determined by $\pGamma_\phi$ and $Q_\psi$, as in 2D flow.

This being the case, here we turn to seeking the {\color{black} asymptotic} force expression with observable physical quantities instead. For this purpose, we first rewrite \er{F1} as
\beqn
  \pF &=& \rho_0 \pU\times \pGamma - \rho_0 \pU\cdot \int_S \pu_\psi \pn \mathrm{d} S +\mu \int_S \po\times \pn \mathrm{d} S, \lb{F2}
\eeqn
where {\color{black} $\pGamma$} is given by \er{Gamma}. Recalling the properties of the {\color{black} transverse} field or the general solution of \er{u-t} \citep[e.g.,][]{Goldstein1931, Garstang1936}, the viscous term in \er{F2} can be omitted and the second term can be replaced by the integral on the wake plane $W$:
\beqn
  - \rho_0 \pU \cdot \int_S \pu_{\psi }\pn \mathrm{d} S &\cong& - \rho_0 \pU \int_W u_{\psi x} \mathrm{d} S. \lb{F3}
\eeqn
To proceed, notice that in the Oseen approximation of unboundedly long steady wake, the variation of flow properties in $x$-direction is much smaller than that in lateral directions, namely $\pat/\pat y, \pat/\pat z \gg \pat/\pat x$ in the {\color{black} wake. Then} the second term of \er{F3} is
\bsubeq\lb{F3-2}
\beqn
  n=2: &\quad& \int_W u_{\psi x} \mathrm{d} S = -\int_W \fr{\pat u_{\psi x}}{\pat z} z \mathrm{d} z \approx - \int_W z \om_y \mathrm{d} S, \\
  n=3: &\quad& \int_W u_{\psi x} \mathrm{d} S = \color{black} -\int_W \fr{\pat u_{\psi x}}{\pat \sigma} \fr{\sigma^2}{2} \mathrm{d} \sigma \mathrm{d} \theta \approx \fr{1}{2} \int_W \sigma \om_\theta \mathrm{d} S,
\eeqn
\esubeq
{\color{black} where $(x, \sigma, \theta)$ is the cylindrical coordinates, and $\sigma \om_\theta = y \om_z -z \om_y$ for $n=3$.} Thus by \er{F3}, \er{F2} {\color{black} reduces} to
\beq\lb{F4}
  \pF \cong \rho_0 \pU\times \pGamma + \rho_0 \pU Q_W,
\eeq
where $\pGamma$ is {\color{black} still} given by \er{Gamma} and
\beq\lb{Q4}
  Q_W \equiv \fr{\rho_0}{n-1} \int_W (z \om_y - y \om_z) \mathrm{d} S,
\eeq
with $n=2,3$ being the space dimensionality. {\color{black} This result} obviously includes and extends the J-F formula \er{F-2D-2}, {\color{black} indicating that} we arrive at a unified far-field asymptotic force formula for both 2D and 3D flows. Thus, we name \er{F4} the \textit{testable unified force formula}. While thus far {\color{black} \er{F2D} and \er{F4} are} derived under the assumed existence of linear far field, {\color{black} this assumption will be rigorously proved in the next two sections from low-speed to supersonic flow.}

\section{Unified force theorem}\label{sec.GG}

In this section we consider the \textit{viscous} flow over a \textit{finite} body, for which \citet[p.~36]{Lagerstrom1964} {\color{black} pointed} out that the linearization is feasible but without proof. The first proof {\color{black} is} given by LZW for two dimensions, {\color{black} using the fundamental solution of} the linear far field {\color{black} of viscous and compressible flow.} This method is now applied to three dimensions, which will lead to a force formula universally true for {\color{black} both 2D and 3D flows}.

\subsection{Fundamental solution method}\label{subsec.fundamental}

For an observer in very far field, a body moving through a fluid appears as a {\it singular point}, {\color{black} its} action on the fluid appears as an {\it impulse force}, {\color{black} and} the far-field disturbance flow is sufficiently weak and {\color{black}  governed} by linearized N-S equations. {\color{black} Then,} to calculate the impulse force there is no need to solve these equations under specified boundary conditions. Rather, it suffices to directly use the {\color{black} corresponding} {\it fundamental solution} in free space. This is the basic idea in the study of linear differential equations, which is called {\it fundamental solution method} and has been successfully demonstrated by LZW for two dimensions.

Following LZW, we introduce (primed) disturbance quantities by
\begin{equation}\label{disturbance}
 \pu=U\pe_x+\pu', \quad
 \rho=\rho_0(1+\rho'),
\end{equation}
then the steady momentum and continuity equations are
\bsubeq\label{final}
\beqn\label{final-1}
  \left(\nu_\theta\bT_\psi-\nu\bT_\phi -U\pat_x \bI \right)\cdot \pu'- c^2\na \rho' &=& -\pf, \\
  \na \cdot \pu' + U\pat_x \rho' &=& 0, \label{final-2}
\eeqn
{\color{black} where}
\begin{eqnarray}
  \bT_\phi =\na \na, \quad
  \bT_\psi = \na\na-\na^2 \bI, \quad
  \bI = \textrm{unit tensor},
\end{eqnarray}
\esubeq
and $\nu = \mu/\rho_0$ and $\nu_\theta=\mu_\theta/\rho_0$ are the constant {\color{black} shear} and longitudinal kinematic viscosities, respectively, $c$ is the speed of sound, and $\pf$ represents an external body force, which in our case is the force exerted to the fluid by the body. In near-field formulation $\pf$ could have a compact distribution in $(\px,t)$-space as used by \citet[p.~51]{Saffman1992}, but below it will be idealized as a $\de$-function of $\px$, i.e.,
\begin{equation}\label{f}
  \pf= -\fr{\delta(\px)}{\rho_0} \pF,
\end{equation}
where $\pF$ is the total force {\color{black} experienced by} the body, {\color{black} $\rho_0$ is the fluid density at infinity, and}
\begin{equation}\label{delta}
  \int \delta(\px) \mathrm{d} \px =1.
\end{equation}

Denote $\bG$ as the fundamental solution of \er{final} for $\pu'$, {\color{black} then}
\beq \lb{u-G}
 \pu'(\px) = \int \bG (\px,\px') \cdot \pf(\px') \mathrm{d} \px'.
\eeq
Since {\color{black} $\pu'$ can be split into a longitudinal part and a transverse} part, see \er{HD}, it can be verified that $\bG$ can also be split into longitudinal and {\color{black} transverse} parts,
\begin{equation}\label{G-g}
  \bG(\px,\px') = \bG_\phi(\px,\px') + \bG_\psi(\px,\px'),
\end{equation}
so that the longitudinal and {\color{black} transverse} velocities defined by \er{HD} now read
\bsubeq\label{uL-uT}
\beqn
  \pu_\phi = \na\phi(\px) &=& \int \bG_\phi (\px,\px') \cdot \pf(\px') \mathrm{d} \px', \label{uL} \\
  \pu_\psi = \na \times\ppsi &=& \int \bG_\psi (\px,\px') \cdot \pf(\px') \mathrm{d} \px'. \label{uT}
\eeqn
\esubeq
Here (see Appendix~\ref{appA}),
\bsubeq\label{G-LT}
\beqn
  \bG_\phi(\px,\px') &=& \fr{1}{4 \pi^2} \int_{-\infty}^\infty\bT_\phi \left[ \fr{e^{i\xi x}}{i\xi U} K_0\left(\sigma \sqrt{ \xi^2 + \fr{i\xi U}{\nu_\theta + \fr{c^2}{i\xi U}}} \right) \right] \mathrm{d} \xi, \lb{G-g-L}\\
  \bG_\psi(\px,\px') &=& -\fr{1}{4 \pi^2} \int_{-\infty}^\infty \bT_\psi \left[ \fr{e^{i\xi x}}{i\xi U} K_0\left(\sigma \sqrt{ \xi^2+\fr{i\xi U}{\nu} } \right) \right] \mathrm{d} \xi \lb{G-g-T}
\eeqn
\esubeq
are the fundamental solutions for the longitudinal and {\color{black} transverse} processes, respectively, and $K_0$ is the modified Bessel function of the second kind.

Compared to dealing with {\color{black} $\pu'$} directly, we find it sometimes more convenient to deal with {\color{black} $\phi$ and $\ppsi$}.
By substituting \er{f} and \er{G-LT} into \er{uL-uT}, we can obtain
\bsubeq\label{phi-psi}
\beqn
 \phi  &=& -\fr{1}{4\pi \rho_0 U} \pF \cdot \na \Phi, \label{phi-1} \\
 \ppsi &=& -\fr{1}{4\pi \rho_0 U} \pF \times \na \Psi, \label{psi-1}
\eeqn
\esubeq
where
\bsubeq\label{phi}
\beqn
  \Phi &\equiv& \fr{1}{\pi} \dashint_{-\infty}^\infty \fr{e^{i\xi x}}{i\xi} K_0\left(\sigma \sqrt{ \xi^2 + \fr{i\xi U}{\nu_\theta + \fr{c^2}{i\xi U}}} \right) \mathrm{d} \xi, \label{phi-2}\\
  \Psi &\equiv& \fr{1}{\pi} \dashint_{-\infty}^\infty \fr{e^{i\xi x}}{i\xi} K_0\left(\sigma \sqrt{ \xi^2+\fr{i\xi U}{\nu} } \right) \mathrm{d} \xi. \label{psi-2}
\eeqn
\esubeq
Here, since the integrals of \er{phi-2} and \er{psi-2} are divergent in general, following \citet{Hadamard1928} the symbol $\dashint$ is used  to denote the finite part of divergent {\color{black} integral.}

\subsection{The transverse far-field}\label{subsec.tran}

By differentiating \er{psi-2} with respect to $x$, there is \citep[p.~722]{Gradshteyn2007}
\beq\lb{Psix-1-3}
  \color{black} \fr{\pat \Psi}{\pat x}= \fr{e^{-k(r-x)}}{r} \equiv \chi,
\eeq
where {\color{black} $k = U/2\nu$ and $r= \sqrt{x^2+y^2+z^2}$.} Then from \er{Psix-1-3} we have
\beq\lb{Psi-2}
  \Psi = \dashint_{-\infty}^x \chi(t,y,z) \mathrm{d} t = \dashint_{r-x}^{\infty} \fr{e^{-kt}}{t} \mathrm{d} t,
\eeq
where we have used the upstream decaying condition $\Psi(-\infty,y,z)=0$. Of course, \er{Psi-2} is independent of Mach number and divergent at the positive $x$-axis $(r-x=0)$, where the singular transverse velocity must be canceled by the longitudinal velocity.

By substituting \er{psi-1} into \er{uT}, the transverse velocity is
\beqn\label{u-T1}
  \pu_\psi &=& \fr{1}{4\pi \rho_0 U}[ \na(\pF\cdot \na\Psi) -\pF\na^2\Psi].
\eeqn
Since
\begin{equation}\label{Psi-4}
  \na^2 \Psi = 2 k \fr{\pat \Psi}{\pat x} = 2k \chi,
\end{equation}
\er{u-T1} yields
\beqn\label{u-T3}
  \pu_\psi &=& \fr{1}{4\pi \rho_0 U} \na(\pF\cdot \na\Psi) +\pv,
\eeqn
where
\begin{equation}\label{u-T-vort}
  \pv \equiv -\fr{1}{4 \pi \mu} \chi \pF
\end{equation}
is the purely rotational velocity.

{\color{black} Then, by \er{u-T3} and \er{u-T-vort} the vorticity is}
\begin{equation}\label{om-T2}
  \po = \na \times \pv = \fr{1}{4\pi \mu} \pF \times \na\chi.
\end{equation}
Because $\na \chi$ is the only vorticity source term in \er{om-T2}, $\chi$ is called the \textit{vorticity potential}, which was first introduced by \citet{Lamb1911} for the linearized far-field of steady axis-symmetrical flow. We now see it does exist for the linearized far-field of {\it any} steady 3D flow. {\color{black} In particular,
\begin{equation}\label{W-om3}
\color{black} \int_W  \chi \mathrm{d} S = \fr{2 \pi}{k} = \rm const.,
\end{equation}
so the wake-plane integral of vorticity vanishes:
\begin{equation}\label{W-om2}
  \int_W \po \mathrm{d} S = \fr{\pF \times \pe_x}{4\pi \mu} \fr{\mathrm{d} }{\mathrm{d} x} \int_W \chi \mathrm{d} S = \bf 0.
\end{equation}

As a check of our algebra, we substitute \er{Psix-1-3} into \er{om-T2} to obtain
\begin{equation}\label{om-T3}
  \po = \fr{k}{4\pi \mu} \fr{e^{-k(r-x)}}{r} \na (r-x) \times \pF + O\left( \fr{e^{-k(r-x)}}{r^2} \right),
\end{equation}
which agrees exactly the asymptotic vorticity expression obtained by \citet{Babenko1973} for incompressible flow \citep[see also][]{Mizumachi1984}. Then, in the $x$-axis there is
\begin{equation}\label{om-T2-x}
  \om_x(x,0,0) = 0, \quad
  \om_y(x,0,0) = -\fr{F_z}{4\pi \mu x^2}, \quad
  \om_z(x,0,0) = \fr{F_y}{4\pi \mu x^2}.
\end{equation}
Thus the lift and side force can also be written as
\begin{equation}\label{om-T2-x2}
  F_y = 4\pi \mu x^2\om_z(x,0,0), \quad
  F_z = - 4\pi \mu x^2\om_y(x,0,0).
\end{equation}
Furthermore,} from \er{u-T-vort} and \er{W-om3} we have
\begin{equation}\label{fi-1}
  \pF \cong -\rho_0 U \int_W \pv \mathrm{d} S,
\end{equation}
which has {\color{black} the same form as \er{F-3D}} since $\pv = \pv_2+\pv_3$. But now it is also valid for compressible flow.
{\color{black} Since} $\pv$ is the purely rotational part of $\pu'$, we may expect that $\pF$ can be solely expressed by the vorticity. This will be discussed {\color{black} thoroughly} in Section \ref{sec.TGG}.

\subsection{Unified force theorem}

With the above preparations, we can now state the following innovative theorem:

{\bf Unified force theorem}. {\sl For an $n$-dimensional steady flow of viscous and compressible fluid over a rigid body, $n=2,3$, the {\color{black} forces} exerted to the body are solely determined by the multi-valueness and singularities of the velocity potential $\phi$ in the circulation $\pGamma_\phi$ and the vortical stream function $\ppsi$ in the inflow $Q_\psi$:}
\beq\lb{f-uni}
  \pF = (n-1)\rho_0 \pU \times \pGamma_\phi + \rho_0 \pU Q_\psi,
\eeq
{\sl where $\pGamma_\phi$ and $Q_\psi$ are given by the first expressions of \er{G-phi} and \er{Q-psi}, respectively, and are independent of the choice of control surface $S$.}

{\it Proof.} First, as remarked previously, either $\pGamma_\phi$ or $Q_\psi$ vanishes due to the generalized Stokes theorem, unless $\phi$ or $\ppsi$ are multi-valued or singular somewhere. This multi-valueness or singularity is independent of the integral surface $S$, and hence so is \er{f-uni}. In fact, $S$ can even be located in the nonlinear near field as long as the definition domain of $\phi$ and $\ppsi$ is properly extended; but the proof of the theorem can be made in the linearized far field where the formal solution \er{phi-psi} is valid.

Then, to prove \er{f-uni}, we only need to show that for $n=3$ there is
\beq\lb{f-uni3D}
  \pF = 2\rho_0 \pU \times \pGamma_\phi + \rho_0 \pU Q_\psi,
\eeq
since then a comparison of \er{f-uni3D} and \er{F2D} implies
\beq\lb{F2D-13}
 \rho_0 \int_S \pn \times \na (\pU \cdot \ppsi) \mathrm{d} S =\rho_0\pU\times \pGamma_\phi,
\eeq
and thus \er{f-uni} follows at once. Note that after the existence of linear far field was proven in {\color{black} Subsections} \ref{subsec.fundamental} and \ref{subsec.tran}, \er{F2D} has become a rigorous result and can be cited here.

To prove \er{f-uni3D}, {\color{black} we substitute \er{u-T3} and \er{phi-1} into \er{HD}, such that}
\beq\lb{HD-2}
  \pu'= -\fr{1}{4\pi \rho_0 U} \na [ \pF \cdot \na (\Phi-\Psi)] + \pv.
\eeq
Since $\pu'$ and $\pv$ must be regular, so must {\color{black} be} $\pF\cdot \na (\Phi-\Psi)$. {\color{black} If we} rewrite \er{phi-1} as
\begin{equation}\label{phi-3}
 \phi = -\fr{1}{4\pi \rho_0 U} [\pF \cdot \na (\Phi-\Psi) +\pF \cdot \na \Psi],
\end{equation}
{\color{black} then the first term makes} no contribution to $\pGamma_\phi$. {\color{black} Thus} \er{G-phi} reduces to (see \ref{appB})
\beq\label{Gamma-2}
 \pGamma_\phi = - \fr{1}{4\pi \rho_0 U} \int_S (\pn\times\na) (\pF \cdot \na \Psi) \mathrm{d} S = \fr{ \pF \times \pe_x}{2\rho_0 U}.
\eeq
Similarly, by substituting \er{psi-1} {\color{black} into} \er{Q-psi}, we find  ({\color{black} see} \ref{appB})
\begin{equation}\label{Q-2}
  Q_\psi = \fr{1}{4\pi \rho_0 U} \int_S (\pn\times\na)\cdot (\pF \times \na \Psi) \mathrm{d} S = \fr{\pF \cdot \pe_x}{\rho_0 U}.
\end{equation}
Obviously, both $\pGamma_\phi$ and $Q_\psi$ are indeed independent of $S$. {\color{black} The proof is thus completed.}

In the above proof we only used the specific behavior of $\Psi$. This approach makes the proof concise and general, but leaves the relevant physical mechanisms of lift and drag obscure. Thus, to explore the underlying physics we still have to directly analyse the disturbance velocity. We do this in the next section.

Before proceeding, we remark that the K-J theorem, the most famous prototype of far-field theory, was basically regarded as the neatest model of engineering theory in the past. Its significance was underestimated due to the ignorance of the normative role of the far-field theories. The reason of this ignorance might come from two facts: (1) There is no far-field theory for inviscid supersonic aerodynamics before and thus its power has not been totally realized; (2) One has now been used to relying heavily on computational fluid dynamics (CFD) but often overlooks its shortages. Therefore, the unified force theorem represents a breakthrough of classic inviscid high-speed aerodynamics, which has set a universal norm for all theories of aerodynamic forces in viscous and compressible steady external flow. To make this normative role indeed take effect in practices, some basic questions are still awaiting to be answered. For example, how does the flow evolve between nonlinear near field and linear far field? Or similarly, how do the near-field theories transform smoothly to the far-field theories? Only with this question being answered can the problem of far-field boundary conditions involved in CFD be completely rationalized. However, these important issues won't be discussed any longer in the present paper.

\section{Testable unified force formula}\label{sec.TGG}

In this section, we shall analyze the longitudinal far-field flow structures, estimate the distances of the linear far field from the body and clarify the far-field behavior of multiple circulations. After that, we further confirm that \er{F4} is the far-field asymptotic approximation of \er{f-uni}, which is expressed solely in terms of observable vorticity field.

\subsection{The longitudinal far-field}

By differentiating \er{phi-2} with respect to $x$, there is
\beq\label{Phix-1}
  \fr{\pat \Phi}{\pat x} = \fr{1}{\pi} \int_{-\infty}^\infty e^{i\xi x} K_0\left(\sigma \sqrt{ A e^{i\theta}} \right) \mathrm{d} \xi,
\eeq
where
\begin{equation}\label{A-theta0}
  A e^{i\theta} \equiv \xi^2 + \fr{i\xi U}{\nu_\theta + \fr{c^2}{i\xi U}} =(1-M^2)\xi^2 + i\fr{\nu_\theta M^4 }{U}\xi^3 + O(\nu_\theta^2), \quad M=\fr{U}{c}.
\end{equation}
Obviously, the longitudinal field described by $\Phi$ depends explicitly on Mach number $M$, as seen from the key factors $1-M^2$ and $M^4$ in \er{A-theta0}.

\subsubsection{Subsonic flow}

For subsonic flow, there is
\begin{equation}\label{theta-sub}
  A \cong \beta^2 \xi^2, \quad
  \theta \cong 0, \quad
  \beta^2 = 1-M^2 >0.
\end{equation}
Then
\beq
  \fr{\pat \Phi}{\pat x} = \fr{1}{r_\beta}, \quad
  r_\beta = \sqrt{ x^2 + \beta^2 (y^2+z^2) }.
\eeq
Thus we have
\beq\lb{Phi-sub-2}
  \Phi = \dashint_{r_\beta-x}^{\infty} \fr{1}{t} \mathrm{d} t = - \ln(r_\beta-x).
\eeq
The longitudinal potential $\phi$ is
\begin{equation}\label{phi-sub1}
 \phi = \fr{1}{4\pi \rho_0 U} \pF \cdot \na \ln(r_\beta-x),
\end{equation}
which is singular at the positive $x$-axis and of which the singular longitudinal velocity can just cancel out that of the transverse field.

\subsubsection{Supersonic flow}

For supersonic flow, there is
\begin{equation}\label{theta-sup}
  A \cong B^2 \xi^2, \quad
  \theta \cong \pi - 2\Lambda \xi, \quad
  B^2 = M^2-1 >0, \quad
  \Lambda = \fr{\nu_\theta M^4}{2 B^2 U} \ll 1.
\end{equation}
Now, we need to find such a viscous solution that is significant only near the Mach cone and decays exponentially elsewhere except in wake region. Firstly, note that
\begin{eqnarray}
  \fr{\pat \Phi}{\pat x} \cong \fr{1}{\pi} \int_{-\infty}^\infty e^{i\xi x} K_0\left( \Lambda B \sigma \xi^2+i B\sigma \xi \right) \mathrm{d} \xi, \label{Phix-sup1}
\end{eqnarray}
then the far-field decaying condition can be ensured since $\Lambda B \sigma \xi^2 \ge 0$, indicating that the viscous effect has a vital role in supersonic flow. However, \er{Phix-sup1} is hard to integrate explicitly. Instead, we consider its approximation at $\sigma =\sqrt{y^2+z^2} \gg 1$ by the asymptotic identity
\begin{eqnarray}
 K_0(z) \cong \sqrt{\fr{\pi}{2z}}e^{-z}, \quad z\rightarrow \infty.
\end{eqnarray}
Then \er{Phix-sup1} reduces to
\beqn\label{Phi-4x}
  \fr{\pat \Phi}{\pat x} \cong \sqrt{\fr{2}{\pi B\sigma}} \int_{0}^\infty \cos\left[ \xi (x-B\sigma) -\fr{\pi}{4} \right] e^{-\Lambda B\sigma \xi^2} \fr{\mathrm{d} \xi}{\sqrt{\xi}}. \label{Phix-sup2}
\eeqn

Although the integral of \er{Phix-sup2} can be worked out explicitly, here we left it out since its form is somewhat complicated and inconvenient to analyze.
Instead, suppose $\pF = D \pe_x$ and substitute it into \er{phi-1}, we have
\beqn
 \phi  &=& -\fr{D}{4\pi \rho_0 U} \fr{\pat \Phi}{\pat x}. \label{phi-sup1}
\eeqn
Then from \er{Phix-sup2} and \er{phi-sup1} and denote $C_D = 2D/\rho_0 U^2$, there is
\begin{eqnarray}
 \fr{1}{U} \fr{\pat \phi}{\pat x} &\cong& \fr{C_D}{8\pi}\sqrt{\fr{2}{\pi B\sigma}} \int_{0}^\infty \sin\left[\xi (x-B\sigma) -\fr{\pi}{4} \right] e^{-\Lambda B\sigma \xi^2} \sqrt{\xi} \mathrm{d} \xi. \label{phix-sup3}
\end{eqnarray}
Since the disturbance must be largest along the Mach cone, there is
\begin{eqnarray}
 \left. \fr{1}{U} \fr{\pat \phi}{\pat x} \right|_{x=B\sigma} \cong -\fr{G(3/4) C_D}{16 \pi^{\fr 3 2} \Lambda^{\fr 3 4} B^{\fr 5 4} \sigma^{\fr 5 4}} . \label{phix-sup4}
\end{eqnarray}
where $G(\cdot)$ is the Gamma function and $G(3/4) = 1.22542\cdots$.

\subsubsection{Sonic flow}

The viscosity is also necessary for sonic flow, where
\begin{equation}\label{A-theta0-sonic}
  \sqrt{ A e^{i\theta} } =  \sqrt{ \fr{|\xi|^3}{Re_\theta}} e^{i \fr{\pi}{4}\, \textrm{sgn\,} \xi}, \quad Re_\theta = \fr{U}{\nu_\theta},
\end{equation}
and
\begin{eqnarray}
  \fr{\pat \Phi}{\pat x} &=& \fr{1}{\pi} \int_{0}^\infty \left[ K_0 \left( \sigma \sqrt{ \fr{ \xi^3}{Re_\theta} } e^{-i \fr{\pi}{4}} \right) e^{-i\xi x} + K_0 \left( \sigma \sqrt{ \fr{ \xi^3}{Re_\theta} } e^{i \fr{\pi}{4}} \right) e^{i\xi x} \right] \mathrm{d} \xi, \label{Phix-son1}
\end{eqnarray}
which for $\sigma \gg  1$ reduces to
\begin{eqnarray}
  \fr{\pat \Phi}{\pat x} &\cong& \sqrt{ \fr{2}{ \pi \sigma} } Re_\theta^{\fr 1 4} \int_0^{\infty} e^{-\sigma \sqrt{ \fr{ \xi^3}{2Re_\theta} } } \cos \left( \fr{\pi}{8} - \xi x + \sigma \sqrt{ \fr{ \xi^3}{2Re_\theta} } \right) \fr{\mathrm{d} \xi}{ \xi^{\fr 3 4}}.
\end{eqnarray}
Furthermore, we have
\begin{eqnarray}
  \fr{\pat^2 \Phi(0,\sigma)}{\pat x^2} \cong \sqrt{ \fr{2}{ 3 \pi} } Re_\theta^{\fr 2 3} \fr{G(5/6)}{ \sigma^\fr{4}{3}}, \lb{phix-son4}
\end{eqnarray}
where $G(5/6) = 1.12879\cdots$.
Thus from \er{phi-sup1} and \er{phix-son4}, there is
\begin{eqnarray}
 \left. \fr{1}{U} \fr{\pat \phi}{\pat x} \right|_{x=0} \cong -\fr{G(5/6) C_D Re_\theta^{\fr 2 3} }{ 4\sqrt{6} \pi^\fr{3}{2} \sigma^\fr{4}{3}}. \label{phix-son5}
\end{eqnarray}

\subsection{Distance of linear far-field from the body}\label{subsec.distance}

We now use the preceding solutions of linear equations to predict how large the \textit{minimum distance} $r_m =\sqrt{x^2_m +\sigma^2_m}$ from the body should be for them to become valid. The estimate is based on a simple requirement that the magnitude of $\pu'$, after being non-dimensionalized, is not larger than unity. For simplicity, we may assume only drag exists. Meanwhile, the corresponding 2D estimates given by LZW are also listed for comparison.

Let the characteristic length scale be unity so that $C_D = 2D/\rho_0 U^2$, $Re =U/\nu$ and $Re_\theta=U/\nu_\theta$. Then, the minimum streamwise distance $x_m$ of the linear far field can be obtained directly from \er{HD-2} by setting $\sigma = 0$ and requiring $|\pv|/U = 1$:
\beq\lb{r-v}
 x_m = O\left( \fr{C_D Re}{8\pi} \right),
\eeq
which is independent of Mach number and fully determined by transverse process. This is comparable with the 2D estimate $r_m =O(C^2_D Re/16\pi)$.

On the other hand, the lateral minimum distance $\sigma_m$ is dominated by longitudinal process, which can be determined from \er{phi-sub1}, \er{phix-sup4} and \er{phix-son5}, respectively:

(1) Subsonic far-field:
\bsubeq\lb{r}
\beq\lb{r-sub}
 \sigma_m = O\left( \sqrt \fr{C_D}{8\pi \beta} \right),
\eeq
which is comparable with the 2D estimate $r_m =O(C_D/4\pi \be)$.

(2) Supersonic far-field:
\beq\lb{r-sup}
 \sigma_m = O\left( \fr{ G(3/4)^{\fr 4 5} B^{\fr 1 5} }{ 2^{\fr{13}{5}} \pi^{\fr 6 5} M^{\fr{12}{5} } } C_D^{\fr 4 5} Re_\theta^{\fr 3 5} \right),
\eeq
which is smaller than 2D estimate $r_m \sim C^2_D Re_\theta$.

(3) Sonic far-field:
\beq\lb{r-sonic}
 \sigma_m = O\left( \fr{ G(5/6)^{\fr 3 4} }{ 96^{\fr 3 8} \pi^{\fr 9 8} } C_D^{\fr 3 4} Re_\theta^{\fr 1 2} \right),
\eeq
\esubeq
which is, remarkably, very much smaller than 2D estimate $r_m \sim C^3_D Re^2_\theta$.

Clearly, in different Mach-number regimes and spatial directions, the dominant linearized far-field dynamic processes and flow structures are vastly distinct, with variance dependence on $C_D$, $Re$ or $Re_\theta$, and $M$. Of these distances $x_m$ is the farthest from the body. Fortunately, this large value is limited in the relatively narrow wake region and will not significantly affect the lift or side force, though it may have strong effect on the drag. It should be stressed that, because $\pu_\phi$ and $\pu_\psi$ are infinite in the positive $x$-axis, one can not determine the location of the transverse and longitudinal fields separately as LZW did for two dimensions.

The above estimates are summarized in Table~\ref{tab.distance}, in order of magnitude. Recall that the friction drag of laminar incompressible flow can be determined as $C_f \sim 1/\sqrt{Re}$, which is the dominate part of $C_D$ in attached low-speed flow. Thus, the location of linear far field of 2D incompressible or subsonic flow is usually independent on $Re$, which, however, is not the case for transonic and supersonic flows or 3D flow. On the other hand, the existence of turbulence may shorten the above estimates. However, this effect only happens in the transverse field or the wake, but has only a negligible effect on the longitudinal field or shock waves. For the latter, let us take the \textit{Concorde} as an example. Assume $M=2$, $Re = 10^9$, $C_D = 0.1$, and the effective chord length $l=10 \, \rm m$, then the linear far field locates as high as $10^4 \, \rm m$, the same order of the flight altitude. This is why the \textit{sonic boom} can be felt even the \textit{Concorde} is at its cruise state.

\begin{table}
\centering
  \begin{tabular}{|c|c|ccc|}  \hline
  & \multicolumn{1}{c|}{Transverse Field, $x_m$} & \multicolumn{3}{c|}{Longitudinal Field, $\sigma_m$} \\
  & $M$-independent & Subsonic & Sonic & Supersonic \\ \hline
  2D & $C_D^2 Re$ & $C_D$ & $C_D^3 Re_\theta^2$ & $C_D^2 Re_\theta$ \\
  3D & $C_D Re$ & $C_D^{ \fr 1 2}$ & $C_D^{ \fr 3 4} Re_\theta^{\fr 1 2}$ & $C_D^{\fr 4 5} Re_\theta^{\fr 3 5}$ \\ \hline
  \end{tabular}
  \caption{Location of linear far field}\label{tab.distance}
\end{table}

\subsection{Multiple circulations}\label{subsec.m-ga}

Similar to the longitudinal circulation $\pGamma_\phi$ given by \er{G-phi} and the total circulation $\pGamma$ given by \er{Gamma}, we can define a transverse circulation (cf.~LZW)
\begin{equation}\label{G-psi}
 \pGamma_\psi \equiv \int_S \pn \times \pu_\psi \mathrm{d} S =\pGamma-\pGamma_\phi.
\end{equation}
Let us now examine the far-field behavior of these three circulations. A substitution of the expression of $\pu'$, \er{HD-2}, into the first expression of \er{Gamma} yields
\beqn\label{Gamma-4}
 \pGamma = \int_S \pn \times \pv \mathrm{d} S,
\eeqn
where $\pv$ is given by \er{u-T-vort}. Let $S$ be a spherical surface with radius $r$ and substitute \er{u-T-vort} into \er{Gamma-4}, there is
\beqn\label{Gamma-15}
 \pGamma = \fr{\pF \times \pe_x}{\rho_0 U} \left( 1-\fr{1}{kr} + e^{-2kr} +\fr{1}{kr} e^{-2kr} \right).
\eeqn
Since $\pGamma_\phi$ is given by \er{Gamma-2}, we obtain
\beqn\label{Gamma-psi3}
 \pGamma_\psi &=& \fr{\pF \times \pe_x}{\rho_0 U} \left( \fr{1}{2} -\fr{1}{kr} + e^{-2kr} +\fr{1}{kr} e^{-2kr} \right), \label{Gamma-psi2}
\eeqn
which is dependent on $r$ or $S$ but independent of Mach number. Unlike two dimensions where $\pGamma_\psi$ decays to zero as $r\rightarrow \infty$, here it converges to the value of $\pGamma_\phi$. This is the reason for the factor {\color{black} $n-1$} in the unified force formula \er{f-uni}.

With these discussions, therefore, the generically non-observable $\pGamma_\phi$ becomes observable when it is used to measure the total vorticity in the whole steady-flow region $V_{\rm st}$:
\beqn\label{Gamma-phi3}
 \lim_{r\rightarrow \infty} \pGamma = 2\pGamma_\phi &=& \int_{V_{\rm st}} \po \mathrm{d} V.
\eeqn

\subsection{Testable unified force formula}\label{subsec.TGG}

Since the {\it integrands} of $\pGamma_\phi$ and $Q_\psi$ are not observable in general cases, we need to find the circumstances in which these integrands can be replaced by physically observable variables. Now the preceding analyses of the flow behaviour have revealed that the possibility lies in the linear far field, of which the existence has been confirmed by the estimates made in Subsection \ref{subsec.distance}. This permits us to give a testable version of the unified force formula \er{f-uni}, which we state first:

{\bf Testable unified force formula}. {\sl For an $n$-dimensional steady flow of viscous and compressible fluid over a rigid body, $n=2,3$, the force exerted on the body is given by \er{F4} with $W$ being the downstream face of the outer boundary of $S$, which is perpendicular to the incoming flow and lies in the linear far field.}

\noindent
Here and below, we will call this result the TUF formula for short. Since its 2D version has been addressed by LZW, we focus on the case $n=3$.

{\bf Remarks}.
1. One of the proofs of the TUF formula is given in Section \ref{sec.far-field}, where use has been made of $\pat/\pat y,\pat/\pat z \gg \pat/\pat x$. In fact, this limit can be removed {\color{black} by transforming} \er{F4} to a form involving wake-plane integrals only. For instance, using the identity \cite[p.~700]{Wu2006}
\begin{equation*}
  \int_V \pf \mathrm{d} V = -\int_V \px (\na \cdot \pf ) \mathrm{d} V + \int_{\pat V} \px (\pn\cdot\pf) \mathrm{d} S,
\end{equation*}
\er{F4} can be cast to
\beq\lb{F5}
  \pF \cong \rho_0\pU\times \int_W \px \om_x \mathrm{d} S -\fr{1}{2} \rho_0 \pU \int_W (y \om_z -z \om_y) \mathrm{d} S.
\eeq
Then, by substituting the expression of vorticity \er{om-T2} into \er{F5} we can directly confirm the validity of \er{F4} free from the aforementioned assumption.

2. The TUF force formulae for $n=2$ and $n=3$ are never equivalent to each other, especially for the drag.

3. Due to its linear dependence on the vorticity, TUF formula \er{F4} is supposed to be valid for statistically stationary flow (such as turbulence).

\section{Physical carrier of singularity: Trailing vortex pair}\label{sec.Physical}

From the TUF formula \er{F4}, we see that the forces are expressible solely in terms of vorticity integrals, from incompressible regime all the way to supersonic regime, no matter what complex processes and structures such as shocks, entropy gradient and curved-shock generated vorticity field may occur. In other words, only vorticity leaves signature in far field since it decays in the wake most slowly and, what is more remarkably, because vorticity is a transverse quantity, \er{F4} is completely independent of Mach number (the specific $M$-dependence of vorticity can only be identified by near-field flow behavior).

This being the case, we may well use the familiar difference in the behaviours of {\it incompressible} vorticity field for $n=2$ and 3 to interpret the distinction in \er{F2D} and \er{F5}. For $n=2$, as discussed in details by LZW, vorticity lines are all straight and along the spanwise direction, making the last term of \er{F2D} vanishes without any singularity. Nevertheless, the flow domain is doubly-connected, permitting multiple values of $\phi$ and $\psi$. This makes it inevitable that a body experiencing a force must have nonzero $[\![\phi]\!]$ and/or $[\![\psi]\!]$, which are responsible for the lift and drag, respectively, and surely independent of the choice of control surface. In contrast, for $n=3$, vorticity lines can be stretched and tilted, and eventually go to far field with $\om_x =-\na^2\psi_x$ being the dominating component there, as indicated by the first term of \er{F5}. This explains why there is an extra term in \er{F2D} for $n=3$: a nonzero $\pU \cdot \ppsi = U \psi_x$ implies a nonzero $\pGamma_\psi$, which gives half of $\pGamma$ (total vorticity in $V_{\rm st}$) at far field with $r\rightarrow \infty$. Now, the flow domain is singly-connected, permitting no multi-valueness of $\phi$ and $\ppsi$. Thus, the only possible mechanism for providing nonzero forces and being independent of the choice of control surface is the singularity of $\phi$ and $\ppsi$, of which the only physical carrier must be the trailing vortex pair for any lifting body.

While a complete analysis of the physical carrier of the singularity for compressible flow is too difficult to be done if not impossible, owing to the $M$-independency of \er{F5} the interpretation first presented by \citet{Goldstein1931} for incompressible is sufficient and worth recapitulating. We do this by a concrete line-vortex doublet model, which is a physical idealization of a vortex filament bounding an infinitely small plane area and of which the strength is equal to the product of the area and the strength of the vortex filament. Suppose $\pF = L \pe_z$ and set $\beta=1$, then \er{phi-sub1} reduces to
\begin{equation}\label{phi-sub2}
 \phi = \fr{L}{4\pi \rho_0 U} \fr{z}{r(r-x)},
\end{equation}
which must be the dominant term in \er{HD} for $y^2 \gg 1$ since the transverse field decays exponentially. Then, in a footnote, \citet{Goldstein1931} asserted that $z/r(r - x)$ gives the potential of a line doublet stretching from the origin along the $x$-axis to plus infinity. Obviously, Goldstein's vortex doublet is precisely the far-field picture of the familiar trailing vortex pair, which can be most intuitively visualized with our fundamental-solution method where the body is shrunk to a singular point.

We now show that this vortex doublet is indeed the only possible source of singularity. Suppose that there is a vortex filament $C$ with strength $\Ga$ and width $b$ such that $\Gamma b$ is fixed as $b\rightarrow 0$. Then the velocity $\pu'$ induced by $C$ is \citep[p.~81]{Wu2006}
\begin{equation}\label{3.32}
 \pu' = \fr{\Gamma}{4\pi} \oint_C \fr{\ptt \times \pr'}{r'^3} \mathrm{d} s = \fr{\Gamma b \pe_z}{4\pi} \fr{1}{r(r-x)}.
\end{equation}
Note that $\pu'$ has only one component $u'_z$ since fluids are entrained from the outside of the doublet forming ``upwash'' and pumped into it forming ``downwash''. But the downwash phenomenon can not be observed from \er{3.32} due to our assumption: $\Gamma \rightarrow \infty$ and $b \rightarrow 0$ but $\Gamma b$ remains fixed. Now consider both $y^2 \gg 1$ and $z^2 \ll 1$, from \er{3.32} we have
\begin{equation}\label{phi-sub3}
 \phi \cong \fr{\Gamma b}{4\pi} \fr{z}{r(r-x)}.
\end{equation}
Recall that in lifting-line theory there is $L \cong \rho_0 U \Ga b$, we see \er{phi-sub3} is identical to \er{phi-sub2} in this special case, which seems to be the first confirmation of the assertion of \citet{Goldstein1931}.

\begin{figure}
  \centering
  \subfigure[$M=0$]{\includegraphics[width=0.45\textwidth]{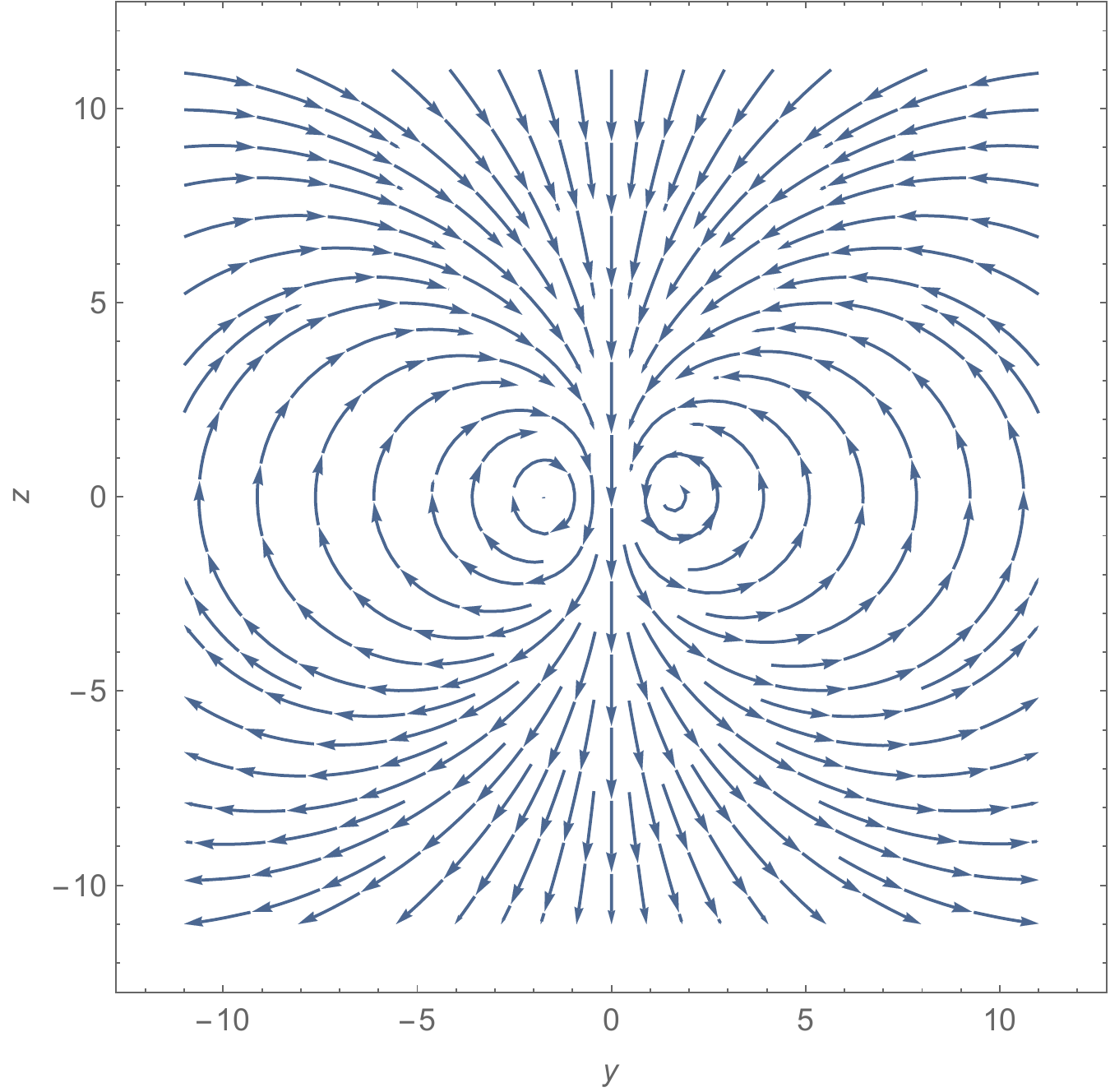}}
  \subfigure[$M=\sqrt 2$]{\includegraphics[width=0.45\textwidth]{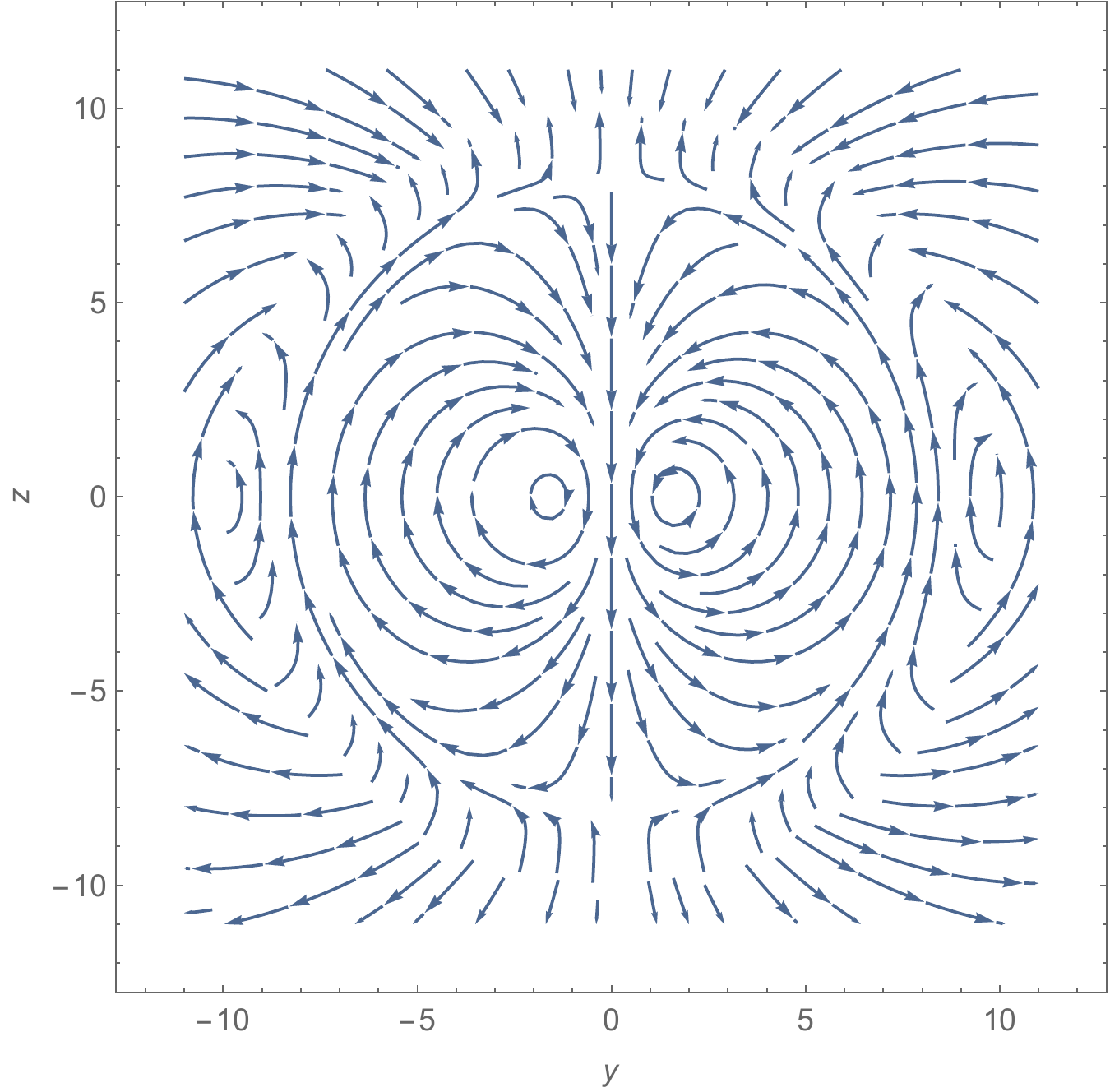}}
  \caption{Vortex-doublet induced cross-flow pattern on a far-field wake plane ($k=x=10$)}
  \label{horse}
\end{figure}

Actually, from the TUF formula \er{F5} we have already known that, the \textit{trailing vortex pair must exist in any steady flow around the lifting-body}, from low-speed to high-speed or even supersonic flight. This observation was first made by \citet{Karman1947} and was further studied by \citet[p.~246]{Liepmann1957} and \citet[Chap.~15]{Ferri1949} with special emphasis on supersonic flow. To visualize the corresponding physical image, Fig.~\ref{horse} shows the vortex-doublet induced cross-flow streamline fields obtained analytically by our theory, namely, \er{Psi-2}, \er{Phi-sub-2} and \er{Phix-sup2}, which are the same as that sketched by \citet{Karman1947} based on physical intuition. The consistence of these two results not only indicates the correctness of our theory, but also once more confirms the fact that the vortex doublet is indeed the physical source of lift.

Note that the above argument is not applicable to non-lifting flow where $\phi$ is regular. However, since $\ppsi$ is still singular there is a drag. This can be seen more clearly by substituting $\pF = D \pe_x$ into \er{psi-1}, which then reduces to
\beq\label{psi-2-2}
 \ppsi = \fr{D}{4\pi \rho_0 U} \fr{e^{-k(r-x)}}{r(r-x)} (0, -z, y).
\eeq
Evidently, the singularity in \er{psi-2-2} comes from the fact that all vorticities are limited to the positive $x$-axis when viewed from far field.

\section{Basic principle to increase the lift-drag ratio}\label{sec.L-D}

As is well-known that, one of the key concerns in aerodynamics is the lift-drag ratio $L/D$. It represents the efficiency of aircraft: the larger $L/D$ implies the better aerodynamic performance and the longer flight range. Evidently, $L/D$ depends on various parameters (e.g., the angle of attack and the flight speed). For typical airplanes, this ratio is $17 \sim 18$ for low-speed and subsonic flights and $10 \sim 12$ for transonic flight. While for the aforementioned \textit{Concorde} at supersonic speed it is $4 \sim 8$.

We now use the TUF formula \er{F4} to give a basic estimate of $L/D$ solely in terms of the far-field vorticity distribution, which is applicable to all aircrafts in steady flight at large range of Mach number. For 2D flow, it can be written as
\beq\lb{L-D}
 \fr{L}{D} = \fr{ \displaystyle \int_{V_{\rm st}} \om \mathrm dV }{ \displaystyle  \int_W z \om \mathrm dz}.
\eeq
This means that, the basic principle to increase $L/D$ should be: {\it to increase the total vorticity in the steady region that is directly related to the lift as large as possible, and to decrease the vorticity moment in the wake plane that is directly related to the drag as small as possible}. One of the typical examples is the supercritical airfoil: through special configuration design, the onset of shock waves in the transonic speed range is significantly delayed and the shock-induced boundary-layer separation is greatly suppressed. As a consequence of the former, the total vorticity in the steady region is larger than that of the common airfoil under the same flow conditions. Meanwhile, as a consequence of the latter, the vorticity moment in the wake plane is also greatly reduced. Thus, $L/D$ is increased.

For 3D flow, the above principle should also be applicable. However, since the vorticity lines or tubes in three dimensions must be stretched and tilted, the corresponding physics behind $L/D$ involves much more abundant mechanisms than that in two dimensions. For instance, from \er{F4} and \er{F5}, $L/D$ can be expressed as
\beq\lb{L-D-3}
 \fr{L}{D} = \fr{ \displaystyle \int_{V_{\rm st}} \om_y \mathrm dV }{ \displaystyle  \fr{1}{2} \int_W (z \om_y - y \om_z) \mathrm dS}  = \fr{ \displaystyle \int_{W} y \om_x \mathrm dS }{ \displaystyle  \fr{1}{2} \int_W (z \om_y - y \om_z) \mathrm dS} .
\eeq
The first equality has the same structure as that in \er{L-D}, confirming that the previous principle indeed applies to 3D flow, and thus needs no further discussion. The second equality is unique for 3D flow and desires more explanations. Take a steady incompressible or subsonic flow over a wing as an example. In the far field, as argued in Section \ref{sec.Physical}, trailing vortex pair forms where $\om_x$ dominates. Although the strength of the trailing vortex is almost constant, the magnitude of $\om_x$ itself decreases along the downstream direction through the viscous diffusion. Due to the divergence-free condition $\na \cdot \po = 0$, therefore, the vortex must be accompanied by relatively small but nonzero $\om_y$ and $\om_z$. Now, suppose that the vorticity field in the wake plane $W$ is similar for different aircrafts under the same flow conditions. Meanwhile, it is also proper to assume that the typical values of $y$ and $z$ in the integrands of \er{L-D-3} are proportional to the characteristic scales of the wingspan $b$ and thickness $h$ of the wing. Thus, we may conclude that: \textit{the larger of the ratio of characteristic wingspan to characteristic thickness of the lifting-body, the better will be its performance}. This should be one of the design principle of aircraft for low-speed or subsonic flight.

It should be pointed out that for transonic and supersonic flights this simple principle of increasing $b/h$ may not work well. This is because vorticities generated by shocks would have a very large spatial extension (especially along the vertical-up direction), making the typical value of $z$ no longer be characterized by $h$ although that of $y$ is still characterized by $b$. For hypersonic flight, however, this principle should resume to be effective as shock waves become very close to the body surface, recovering the characteristic role of $h$.

\section{Conclusions}\label{sec.Conclusions}

In this theoretical paper, we have studied the forces experienced by a body in a steady flow of viscous and compressible fluid. The major findings are summarized as follows:

1. A unified force theorem has been obtained and proved to hold universally for both two and three dimensions. It states that the forces exerted on the body are unified determined by the vector circulation $\pGamma_\phi$ of longitudinal velocity and scalar inflow $Q_\psi$ of transverse velocity, both being independent of the choice of control surface used to calculate the circulation and inflow.

2. The far-field asymptotic form of the exact unified force formula has also been obtained, which is solely expressed by vorticity integrals and valid if the control surface lies in the linear far field. Its form is also independent of the Reynolds number and Mach number. This result is a reflection of the inherent flow physics: no matter how many interacting processes could appear in the nonlinear near-field flow, only the vorticity field has the farthest downstream extension and leaves signature in the far field.

3. The unified force formula and its far-field asymptotics contain explicitly the space dimensionality, so the expressions of forces are never the same for two and three dimensions. This fact is a result of the intrinsic difference of flow patterns in two and three dimensions. Unlike two dimensions where the forces come solely from the multi-valueness of velocity potentials $\phi$ and $\ppsi$ in doubly-connected flow domain, now in three dimensions they come solely from the singularity of $\phi$ and $\ppsi$ in singly-connected flow domain. As the body shrinks to a point in the far field, the trailing vortex pair degenerates to a line-vortex doublet of vanishingly width, which is the only physical carrier of singularity in both incompressible and compressible flows.

\begin{acknowledgments}
This work was partially supported by National Natural Science Foundation of China (Grant No. 10921202, 11221062, 11521091, 11472016). The authors wish to thank Prof. Yi-Peng Shi, Dr. Jin-Yang Zhu, Messrs. Shu-Fan Zou and An-Kang Gao for very valuable discussions. The authors also appreciate the valuable comments of the anonymous referees. The first author gratefully acknowledges the support of the Boya Postdoctoral Fellowship.
\end{acknowledgments}

\appendix

\section{Fundamental solution of the linearized Navier-Stokes equations}\label{appA}

Here we give the derivation of the fundamental solution of \er{final} for $\pu'$ in three dimensions. Denote the Fourier transform and inverse transform of \er{final} in $x$-direction as
\begin{equation}\label{d79a}
  \widetilde f(\xi,y,z) = \int_{-\infty}^\infty e^{-i\xi x} f(x,y,z) \mathrm{d} x, \quad
  f(x,y,z) = \fr{1}{2\pi} \int_{-\infty}^\infty e^{i\xi x} \widetilde f(\xi,y,z) \mathrm{d} \xi.
\end{equation}
Now equations in \er{final} are transformed to (upon eliminating $\widetilde \rho'$)
\begin{equation}\label{d82}
  \left( a\bM_1 -b\bM_2 - k^2 \bI \right) \cdot \widetilde \pu' = - \widetilde \pf,
\end{equation}
where
\bsubeq
\beq\lb{abk}
 a=\nu_\theta + \displaystyle \fr{c^2}{i \xi U} , \quad
 b=\nu, \quad k^2 =i \xi U,
\eeq
and
\beq\lb{T-LT}
 \bM_1 = \widetilde\na \widetilde\na,  \quad
 \bM_2 = \widetilde\na \widetilde\na - \widetilde\na^2 \bI, \quad
 \widetilde \na = (i\xi, \pat_y, \pat_z).
\eeq
\esubeq
The fundamental solution of \er{d82} can be obtained by the theorem \citep[pp.~172-175]{Lagerstrom1949}:

\textbf{Theorem}. {\sl If $\bM_1$ and $\bM_2$ are two linear differential matrix operators such that}
\begin{equation}\label{d43a}
  \bM_1 \cdot \bM_2 = \bM_2 \cdot \bM_1 = \mathbf{0}, \quad
  \bM_1 - \bM_2 = L \bI,
\end{equation}
{\sl where $\bI$ is the unit matrix and $L$ is a scalar linear differential operator, then the fundamental solution} $\bG(\px, \px')$ {\sl of \er{d82} is given by}
\begin{equation}\label{d44}
  \bG(\px,\px') = \fr{1}{k^2} \left( \bM_1 g_{\sqrt{\fr{k^2}{a}}} - \bM_2 g_{\sqrt{\fr{k^2}{b}}} \right),
\end{equation}
{\sl where} $g_h(\px,\px')$ {\sl is the fundamental solution of the scalar operator $L-h^2$ with $h$ denotes either $\sqrt{k^2/a}$ or $\sqrt{k^2/b}$. }

Now, since $L = \pat^2_y+\pat^2_z - \xi^2$, of which the fundamental solution $g_h$ with far-field decaying condition is \citep[p.~178]{Lagerstrom1949} \begin{equation}\label{d59}
  g_h = \fr{ 1 }{2\pi}K_0( \sigma\sqrt{h^2+\xi^2}), \quad \sigma=\sqrt{y^2+z^2},
\end{equation}
where $K_0$ is the modified Bessel function of the second kind. Then the fundamental solution $\widetilde \bG$ of \er{d82} comes from \er{d44} and \er{d59} directly,
\begin{eqnarray}
 \widetilde \bG = \fr{1}{2\pi k^2} \left[ \bM_1 K_0 \left( \sigma \sqrt{\fr{k^2}{a}+\xi^2} \right) - \bM_2 K_0 \left( \sigma \sqrt{\fr{k^2}{b}+\xi^2} \right) \right]. \lb{d90}
\end{eqnarray}
Transforming back to the physical space, we obtain
\begin{equation}\label{G}
  \bG = \fr{1}{2\pi} \int_{-\infty}^\infty \fr{ e^{i\xi x} }{2\pi k^2} \left[ \bM_1 K_0 \left( \sigma \sqrt{\fr{k^2}{a}+\xi^2} \right) - \bM_2 K_0 \left( \sigma \sqrt{\fr{k^2}{b}+\xi^2} \right) \right] \mathrm{d} \xi,
\end{equation}
with $a, b, k^2$ and $\bM_1, \bM_2$ being given by \er{abk} and \er{T-LT}, respectively.

\section{The calculations of circulation and inflow}\label{appB}

The circulation $\pGamma_\phi$ and the inflow $Q_\psi$ are given by \er{Gamma-2} and \er{Q-2}, respectively,
\bsubeq\label{Gamma-Q-7}
\begin{eqnarray}
  \pGamma_\phi &=& - \fr{1}{4\pi \rho_0 U} \int_S (\pn\times\na) (\pF \cdot \na \Psi) \mathrm{d}S, \label{Gamma-7} \\
  Q_\psi &=& \fr{1}{4\pi \rho_0 U} \int_S (\pn\times\na)\cdot (\pF \times \na \Psi) \mathrm{d} S, \label{Q-7}
\end{eqnarray}
\esubeq
where
\begin{eqnarray}\label{na-Psi-3}
  \na \Psi = \chi \left(1, -\fr{y}{r-x}, -\fr{z}{r-x} \right), \quad
  \chi = \fr{e^{-k(r-x)}}{r}.
\end{eqnarray}
Due to the factor $e^{-k(r-x)}$ in \er{na-Psi-3}, \er{Q-7} can be reduced to a wake-plane integral,
\begin{eqnarray}
  Q_\psi &\cong& \fr{1}{4\pi \rho_0 U} \int_W (- \pe_y \pat_z + \pe_z \pat_y) \cdot (\pF \times \na \Psi) \mathrm{d} S. \label{Q-8}
\end{eqnarray}

Since \er{Gamma-Q-7} depends linearly on $\pF$, we can estimate its results by assigning $\pF$ with specific values. Suppose $\pF = D \pe_x$, then $\pGamma_\phi \equiv \bf 0$ as $\chi$ is regular while \er{Q-8} reduces to
\beqn
 Q_\psi &=& - \fr{D}{4\pi \rho_0 U} \int_S \left[ \fr{\pat}{\pat y} \fr{y e^{-k(r-x)}}{r(r-x)} + \fr{\pat}{\pat z} \fr{z e^{-k(r-x)}}{r(r-x)} \right] \mathrm{d} S \nonumber \\
 &\cong& \fr{D}{4\pi \rho_0 U} \int_W \fr{kr^2 + krx + x}{r^3} e^{-k(r-x)} \mathrm{d} S =\fr{D}{\rho_0 U}.
\eeqn
Due to the symmetry of $y$ and $z$ in \er{na-Psi-3}, for the lift or side force case we only need to consider $\pF = L \pe_z$ in \er{Gamma-7} or \er{Q-8}. In this case, \er{Q-8} reduces to
\beqn
 Q_\psi &\cong& \fr{L}{4\pi \rho_0 U} \int_W \fr{\pat}{\pat z} \fr{e^{-k(r-x)}}{r} \mathrm{d} y \mathrm{d} z =0.
\eeqn
However, \er{Gamma-7} needs more algebraic operations, which can be simplified by letting $S$ be a sphere surface with $\pn = \pe_r$. Then, \er{Gamma-7} reduces to
\begin{eqnarray}
 \pGamma_\phi = (0, \Gamma_{\phi y}, \Gamma_{\phi z}), \label{Gamma-5}
\end{eqnarray}
where
\beqn
 \Gamma_{\phi y} &=& \fr{L}{4\pi \rho_0 U} \int_S (\pe_y \times \pe_r) \cdot \na \left[ \fr{ze^{-k(r-x)}}{r(r-x)} \right] \mathrm{d} S \nonumber \\
 &=& \fr{L}{4\pi\rho_0 U} \int_S \fr{x^2+z^2 - r x + kz^2(r-x)}{r^2 (r-x)^2} e^{-k(r-x)} \mathrm{d} S\nonumber \\
 &=& \fr{L}{4\pi \rho_0 U} \int_0^{\pi} \int_0^{2\pi} \fr{\cos^2\theta - \cos\theta + \sin^2\theta \sin^2\varphi [ 1 + kr (1-\cos\theta)]}{(1-\cos\theta)^2 e^{kr(1-\cos\theta)}} \sin\theta \mathrm{d} \theta \mathrm{d} \varphi \nonumber \\
 &=& \fr{L}{4 \rho_0 U} \int_{-1}^{1} \fr{2t^2 + (1-t^2) - 2t + kr (1-t^2)(1-t)}{(1-t)^2 e^{kr(1-t)}} \mathrm{d} t = \fr{L}{2 \rho_0 U},
\eeqn
and
\beqn
 \Gamma_{\phi z} &=& \fr{L}{4\pi \rho_0 U} \int_S (\pe_z \times \pe_r) \cdot \na \left[ \fr{ze^{-k(r-x)}}{r(r-x)} \right] \mathrm{d} S \nonumber \\
 &=& -\fr{L}{4\pi \rho_0 U} \int_S \fr{1+k(r-x)}{r^2 (r-x)^2} y z e^{-k(r-x)} \mathrm{d} S =0.
\eeqn

\bibliography{mybibfile}

\end{document}